\documentclass[sigconf,nonacm, natbib=false]{acmart}
\AtBeginDocument{%
  }

\usepackage{multirow}
\usepackage{subcaption}
\usepackage{rotating}
\usepackage{microtype}
\usepackage{tikz}
\usepackage{colortbl}

\RequirePackage[
  datamodel=acmdatamodel,
  style=acmnumeric,
  ]{biblatex}

\begin{document}

\title{Barlow Twins for Sequential Recommendation}

\author{Ivan Razvorotnev}
\email{ivan.razvorotnev@skoltech.ru}
\affiliation{%
  \institution{Skoltech, Higher School of Economics}
  \city{Moscow}
  \country{Russia}
}

\author{Marina Munkhoeva}
%\email{munkhoeva@airi.net}
\affiliation{%
  \institution{AIRI}
  \city{Moscow}
  \country{Russia}
}

\author{Evgeny Frolov}
%\email{frolov@airi.net}
\affiliation{%
  \institution{AIRI, HSE University}
  \city{Moscow}
  \country{Russia}
}

\renewcommand{\shortauthors}{Razvorotnev et al.}

\begin{abstract}

Sequential recommendation models must navigate sparse interaction data, popularity bias, and conflicting objectives like accuracy versus diversity. While recent contrastive self-supervised learning (SSL) methods offer improved accuracy, they come with trade-offs: large batch requirements, reliance on handcrafted augmentations, and negative sampling that can reinforce popularity bias.
In this paper, we introduce BT-SR, a novel non-contrastive SSL framework that integrates the Barlow Twins redundancy-reduction principle into a Transformer-based next-item recommender. BT-SR learns embeddings that align users with similar short-term behaviors while preserving long-term distinctions—without requiring negative sampling or artificial perturbations. This structure-sensitive alignment allows BT-SR to more effectively recognize emerging user intent and mitigate the influence of noisy historical context.
Our experiments on five public benchmarks demonstrate that BT-SR consistently improves next-item prediction accuracy and significantly enhances long-tail item coverage and recommendation calibration. Crucially, we show that a single hyperparameter can control the accuracy-diversity trade-off, enabling practitioners to adapt recommendations to specific application needs. 
% These findings highlight non-contrastive learning as a powerful, efficient, and more controllable alternative in sequential recommendation.

% Non-contrastive learning (NCL) methods have recently advanced many areas of machine learning, often outperforming their contrastive counterparts.  Among these, Barlow Twins excels at producing embeddings whose dimensions capture distinct, non-redundant features while remaining robust to perturbations.  Despite these benefits, NCL remains almost entirely unexplored within the recommender systems domain.  In this paper, we bridge that gap by adapting Barlow Twins to sequential recommendation and examining its effects on model behavior.  Specifically, we build on the Transformer-based next-item predictor, augmenting its standard cross-entropy objective with a Barlow Twins redundancy-reduction term.  Our resulting Barlow Twins for Sequential Recommendation (BT-SR) model not only outperforms strong baselines in accuracy, but also mitigates popularity bias and produces more confident, well-calibrated recommendations.

\end{abstract}




% \received{29 april 2025}

\maketitle

\section{Introduction}
\label{sec:intro}

Sequential recommendation models have demonstrated impressive performance by capturing the rich temporal dynamics inherent in user–item interactions across many recommender system domains and applications. Transformer-based~\cite{DBLP:journals/corr/VaswaniSPUJGKP17} architectures remain the state of the art~\cite{Mezentsev_2024, DBLP:conf/ecir/PetrovM24, DBLP:journals/corr/abs-1808-09781} thanks to their ability to model long-range dependencies and complex sequential patterns in users' consumption histories. However, the extreme sparsity typical for such behavioral data often hampers these models’ capacity to learn robust user representations from limited interaction information.

Real‐world sequential recommender systems must contend with several intertwined problems. First, \emph{popularity bias} skews recommendations toward a small subset of ``short-head'' items. It  marginalizes long‐tail content and reduces overall catalog coverage. Second, \emph{metric trade‐offs} force practitioners to balance conflicting objectives such as precision versus recall, short‐list accuracy versus long‐list engagement, or accuracy versus novelty, therefore making one‐size‐fits‐all solutions impractical. Third, achieving true \emph{personalization} requires reconciling short‐term session dynamics with stable long‐term user preferences, a tension that often leads to either myopic or overly generic suggestions. Finally, \emph{diversification} is essential to avoid repetitive item sequences and to expose users to a broader range of content, yet naive diversification can compromise relevance.

To address the sparsity issue, recent work has integrated contrastive learning (CL), a class of self-supervised learning (SSL) methods, into sequential recommenders~\cite{DBLP:journals/corr/abs-2010-14395, DBLP:journals/corr/abs-2110-05730, du2022contrastivelearningbidirectionaltransformers, liu2021contrastiveselfsupervisedsequentialrecommendation}. By pulling together augmented views of the same user sequence and pushing apart different sequences, these methods can substantially boost recommendation accuracy. Yet contrastive approaches introduce new challenges: (1) \emph{popularity‐based negative sampling} can exacerbate popularity bias~\cite{cai2024popularityawarealignmentcontrastmitigating}, reducing novelty and catalog coverage and reinforcing a feedback loop that disproportionately favors already-popular items; (2) acquiring high‐quality negatives demands \emph{large batch sizes}, inflating memory and computational overhead~\cite{DBLP:journals/corr/abs-2010-04592, kulatilleke2022efficientblockcontrastivelearning}; and (3) \emph{representation collapse}~\cite{guo2024embeddingcollapsescalingrecommendation}, where embeddings become overly similar, diminishing the model’s expressivity. 

Conversely, \emph{non-contrastive self-supervised methods} have recently achieved remarkable results in computer vision~\cite{chattopadhyay2023evaluationnoncontrastiveselfsupervisedlearning} and natural language processing~\cite{shiao2023linkpredictionnoncontrastivelearning} without relying on negative samples~\cite{zhuo2023unifiedtheoreticalunderstandingnoncontrastive}. Among these, the Barlow Twins framework~\cite{pmlr-v139-zbontar21a} stands out: it jointly enforces invariance to perturbations and reduces redundancy across embedding dimensions. 

Applied to sequential recommendation, Barlow Twins can help learn diverse, informative user representations with far lower computational cost and without sampling negatives. By explicitly minimizing redundancy across embedding dimensions, the learned representations can better capture \emph{varied aspects} of user behavior, which is critical for increasing recommendation diversity and long-tail coverage. However, the potential of this approach has not been fully studied in prior recommender system literature, which sets the basis of the current work.

This research investigates the application of non-contrastive learning principles to enhance sequential recommender systems. Specifically, we address the following research questions:
% \textbf{RQ1:} 
(i) can non-contrastive learning produce higher-quality user embedding spaces than contrastive learning within the recommendation domain?
% \textbf{RQ2:} 
(ii) does non-contrastive learning effectively mitigate popularity bias in recommendations?
% \textbf{RQ3:}
(iii) how sensitive are the resulting recommendations to the hyperparameters of the non-contrastive loss function?

This paper presents BT-SR (Barlow Twins for Sequential Recommendation), the first framework to systematically adapt the Barlow Twins redundancy‐reduction objective to next‐item prediction. Our approach augments a standard Transformer‐based recommender with an auxiliary Barlow Twins loss that (i) enforces invariance under sequence augmentations and (ii) discourages redundant features in the embedding space. 
By tuning the relative weight of this loss, BT-SR enables practitioners to shift recommendation behavior—balancing precision on head items against long-tail coverage, and short-list accuracy against long-range engagement—without sacrificing overall accuracy.
% By tuning the relative weight of this loss, BT-SR allows practitioners to navigate trade-offs between head and tail items coverage as well as between short‐list and long‐list recommendation performance. % 

Our main contributions are:
\begin{itemize}
  \item \textbf{A multi‐task Transformer architecture:} We seamlessly integrate the Barlow Twins redundancy‐reduction objective with next‐item prediction in a unified training loop.
  \item \textbf{Controllable recommendation behavior:} We show how a single hyperparameter in BT-SR steers the model along the spectrum of popularity bias—trading off precision on head items versus enhanced coverage of long-tail items, and balancing short‐list versus long‐list metrics. This controllability enables system designers to align recommendation behavior with application-specific goals.
  \item \textbf{Reduced popularity bias and improved calibration:} Through extensive experiments, we demonstrate that BT-SR not only mitigates over‐recommendation of popular items but also produces more confident and better‐calibrated predictions.
\end{itemize}

Extensive empirical studies on several benchmark datasets confirm that BT-SR consistently outperforms state‐of‐the‐art sequential recommenders and contrastive‐learning alternatives.

% To address these questions, we propose \textbf{BT-SR} (Barlow Twins for Sequential Recommendation), a novel architecture that learns user embeddings directly from augmented interaction sequences using the Barlow Twins loss. Through extensive experiments on multiple benchmark datasets, we demonstrate that BT-SR outperforms state-of-the-art sequential recommendation models and contrastive learning baselines.

%\newpage

\section{Related Works}

\subsection{Transformers and Self-Supervised Learning in Sequential Recommendation}

SASRec~\cite{DBLP:journals/corr/abs-1808-09781} and BERT4Rec~\cite{sun2019bert4recsequentialrecommendationbidirectional} pioneered the application of Transformer-based architectures to sequential recommendation. These models have since become the de facto standard due to their capacity to capture long-range dependencies and encode rich user intent representations. Subsequent research has focused on refining their architectural and training components, including attention mechanisms, positional encodings, and optimization objectives~\cite{DBLP:conf/ecir/PetrovM24}.

To mitigate data sparsity and enhance representation learning, recent advances have integrated self-supervised learning (SSL) into Transformer-based recommenders. A predominant line of work follows the contrastive learning (CL) paradigm, wherein positive views are generated via data augmentations and pulled closer in the embedding space, while negatives are repelled. CL4Rec~\cite{DBLP:journals/corr/abs-2010-14395} introduced four canonical augmentation strategies—cropping, reordering, masking, and substitution—whereas DuoRec~\cite{DBLP:journals/corr/abs-2110-05730} proposed target-aware augmentations to preserve user intent and sequence semantics. Further refinements, such as CBiT~\cite{du2022contrastivelearningbidirectionaltransformers} and SCL~\cite{conf/ecir/ShiWL24}, have incorporated bidirectional attention mechanisms and embedding uniformity regularization to stabilize optimization and improve generalization.

Despite their empirical success, CL-based frameworks suffer from several intrinsic limitations. First, their reliance on negative sampling incurs additional computational cost and often introduces popularity bias~\cite{cai2024popularityawarealignmentcontrastmitigating}. Second, their performance exhibits high sensitivity to batch size and temperature hyperparameters~\cite{DBLP:journals/corr/abs-2010-04592}, making them less robust in large-scale or imbalanced recommendation settings. These challenges have motivated growing interest in non-contrastive self-supervised paradigms that eliminate the dependency on explicit negatives.

\subsection{Next-item Prediction Objectives}

To establish a principled baseline for our next-item recommendation model, we build upon the SASRec architecture and systematically compare three loss formulations—binary cross-entropy (BCE), full softmax cross-entropy (CE), and scalable cross-entropy (SCE). Each objective exhibits distinct trade-offs between predictive accuracy, optimization stability, and computational scalability. Below, we formalize these objectives and discuss their respective characteristics.

We first formulate next-item recommendation as a binary classification problem and optimize the binary cross-entropy loss:
\begin{equation}
\mathcal{L}_{\mathrm{BCE}}(\theta)
=
-\sum_{u\in\mathcal{U}}
\Biggl[
\sum_{t=2}^{n}
\log p_{\theta}(i^u_t, t)
+
\sum_{j\in S_{u,t}}
\log\bigl(1 - p_{\theta}(j, t)\bigr)
\Biggr],
\label{eq:bce_loss}
\end{equation}
where
\[
p_{\theta}(i,t) = \sigma\bigl(r_{\theta}(i,t)\bigr),
\quad
\sigma(x) = \frac{1}{1 + e^{-x}},
\]
and \(S_{u,t}\subset\mathcal{I}\) denotes a set of negative samples (items not yet interacted with by \(u\) before step \(t\)). In practice, one or a few negatives are drawn per step to approximate the full objective.

Recent work has demonstrated that replacing SASRec’s binary cross-entropy with the full softmax-based cross-entropy substantially improves next-item prediction accuracy by framing the task as a multinomial logistic regression problem. Given the scoring function \(r_\theta(i,t)\), the probability of recommending item \(i\) at step \(t\) is expressed as
\begin{equation}
p^c_{\theta}(i,t) \;=\;\mathrm{softmax}\bigl(r_\theta(i,t)\bigr)
\quad\bigl(\mathrm{over\ all\ }i\in\mathcal{I}\bigr),
\end{equation}
yielding the cross-entropy objective
\begin{equation}
\mathcal{L}_{\mathrm{CE}}(\theta)
\;=\;
- \sum_{u\in\mathcal{U}}
  \sum_{t=2}^{n}
    \log p^c_{\theta}\bigl(i^u_t,\,t\bigr).
\end{equation}
While this formulation achieves superior predictive performance, computing the full softmax over the entire item catalog incurs \(O(|\mathcal{U}|\cdot N)\) time and space complexity, which becomes intractable for large-scale recommendation systems. Sampling-based approximations or partial normalization strategies thus remain essential for scalability.

To address this computational bottleneck, \cite{Mezentsev_2024} proposed the Scalable Cross-Entropy (SCE)—a state-of-the-art approximation that replaces the full normalization term with a dynamically sampled candidate set. Let \(\mathcal{C}_{u,t}\subset\mathcal{I}\) denote a subset containing the true next item \(i^u_t\) and \(K\) negative samples drawn uniformly or from a learned proposal distribution. Then,
\[
p^{s}_{\theta}(i,t)
=\frac{\exp\bigl(r_{\theta}(i,t)\bigr)}{\sum_{j\in\mathcal{C}_{u,t}}\exp\bigl(r_{\theta}(j,t)\bigr)}
\quad\bigl(i\in\mathcal{C}_{u,t}\bigr),
\]
and the corresponding sampled cross-entropy loss is given by
\begin{equation}
\mathcal{L}_{\mathrm{SCE}}(\theta)
=
- \sum_{u\in\mathcal{U}}
  \sum_{t=2}^{n}
    \log p^{s}_{\theta}\bigl(i^u_t,\,t\bigr).
\end{equation}
By constraining the normalization to the sampled candidate set \(\mathcal{C}_{u,t}\), SCE reduces both time and memory complexity to \(O(|\mathcal{U}|\cdot K)\) per update, thereby enabling efficient training on catalogs with millions of items. In this study, we combine the Barlow Twins regularization term with each of the BCE, CE, and SCE objectives to assess its influence across different backbone–loss configurations.

\subsection{Non-Contrastive Learning for Representation Learning}

Non-contrastive self-supervised learning (NCL) aims to learn invariant and discriminative representations without relying on explicit negative samples. Representative NCL frameworks include BYOL~\cite{grill2020bootstrap}, SimSiam~\cite{chen2021exploring}, VICReg~\cite{bardes2022vicreg}, and Barlow Twins~\cite{zbontar2021barlow}. These methods employ architectural asymmetries—such as stop-gradient operations, momentum encoders, or prediction heads—and objective-specific regularizations, including redundancy reduction and variance normalization, to avert representational collapse.

Although these paradigms have achieved remarkable success in computer vision and have recently been extended to NLP~\cite{shiao2023linkpredictionnoncontrastivelearning}, their application to recommender systems, and particularly to sequential models, remains relatively underexplored. Existing efforts primarily adopt an offline pretraining–fine-tuning pipeline: for instance, CLUE~\cite{Cheng2021LearningTU} adapts BYOL for collaborative filtering through offline self-supervised pretraining, and SelfCF~\cite{liu2021contrastiveselfsupervisedsequentialrecommendation} integrates self-supervision into non-sequential matrix factorization models.

The most relevant prior work, NCL-SR~\cite{zeng2025a}, extends non-contrastive learning to sequential recommendation; however, it relies on external side information to construct augmented views and lacks a systematic analysis of model controllability and representation behavior. In contrast, our BT-SR framework introduces a unified multi-task training scheme that integrates the Barlow Twins objective directly with the next-item prediction loss, enabling controllable redundancy reduction during training.

Furthermore, recent work~\cite{google25} employed the Barlow Twins objective for SSL-based pretraining followed by fine-tuning on downstream recommendation tasks, including next-item prediction. We argue that such a two-stage pipeline is suboptimal for the recommendation domain: (i) SSL pretraining is substantially more time-consuming than joint multi-task optimization, and (ii) the multitask setup, wherein the Barlow Twins loss functions as a regularizer, allows for explicit control over recommendation diversity and behavioral bias through a single hyperparameter. This property is central to our contribution and will be further discussed in subsequent sections.

\section{Proposed Approach}

The challenges outlined in Section~\ref{sec:intro}—such as popularity bias, personalization trade-offs, and embedding sparsity—are rooted in how user interaction sequences are represented. Transformer-based sequential recommenders rely on fixed-length embeddings to summarize user histories, but these embeddings often struggle to balance two critical goals: capturing shared short-term intent and preserving user-specific long-term preferences. This imbalance leads to brittle recommendations, especially in sparse or skewed data settings.

In this section, we introduce \textbf{BT-SR}, a non-contrastive learning framework that enhances sequence embeddings by reducing feature redundancy and promoting invariance under meaningful augmentations. We first formalize the problem setting and describe the next-item prediction objective used as a base task (Section~\ref{sec:prelims}). Next, we introduce \emph{our supervised augmentation strategy, which defines semantically meaningful views of user sequences} (Section~\ref{subsec:augs}). Finally, we present the Barlow Twins loss as an auxiliary objective that encourages redundancy reduction and representation alignment across augmented views (Section~\ref{sec:bt}).

\subsection{Problem Setup and Preliminaries}
\label{sec:prelims}
We consider a standard sequential recommendation setting where the goal is to predict the next item a user will interact with, given their past behavior.
Let $\mathcal{U}$ denote the set of users and $\mathcal{I} = \{i_1, \dots, i_N\}$ the catalog of items. Each user $u \in \mathcal{U}$ has a chronological interaction sequence:
\[
S_u = \bigl(i^u_{\pi^u_1}, i^u_{\pi^u_2}, \dots, i^u_{\pi^u_{|S_u|}}\bigr),
\]
where $\pi^u_t$ indexes interactions by timestamp. The goal of sequential recommendation is to learn a function that, given the $n$ most recent items from $S_u$, accurately predicts the next item. Formally, the model is trained to maximize the likelihood
\begin{equation}
P_{\theta}\bigl(i^u_t \mid i^u_{t-n}, \dots, i^u_{t-1}\bigr),
\label{eq:next_item_likelihood}
\end{equation}
where $\theta$ denotes model parameters.

In practice, this is implemented by encoding the user history at moment $t$ in the form of the item sequence \mbox{$s_u(t) = [i^u_{t-n}, \dots, i^u_{t-1}]$} into a representation $z_u(t) = f_\theta(s_u(t))$ using a Transformer encoder. The next-item prediction task is then formulated as a classification problem based on the relevance scores over the entire item catalog. The scoring function is typically defined in a matrix-factorization style as a scalar product between an item embedding $e_i$ from the catalog and the current sequence state: $r_\theta(i, t) = e_i^\top z_u(t)$. We denote the classification loss as \(\mathcal{L}_{\text{pred}}\).

As baselines, we use SASRec variants trained with binary cross-entropy (BCE), full softmax cross-entropy (CE), and scalable sampled softmax (SCE), described in detail in Section~\ref{subsec:baselines}.

While effective for optimizing next-item accuracy, these objectives do not explicitly encourage structural alignment across semantically similar user sequences. In what follows, we introduce an augmentation scheme designed to reveal such alignment, followed by an auxiliary redundancy-reduction objective that strengthens the representational structure of the embedding space.

\subsection{Designing Supervised Augmentations}
\label{subsec:augs}
To induce semantically meaningful alignment between users, we design a supervised augmentation scheme guided by the next-item label. Inspired by prior work on label-guided contrastive learning~\cite{DBLP:journals/corr/abs-2110-05730}, we construct augmentations based on shared recent behavior. Specifically, given an anchor sequence $S_u = [i^u_1, \dots, i^u_t]$ ending with target item $i^u_t$, we uniformly sample another sequence $S_{u'}$ from the training set whose final item is also $i^u_t$. These two sequences, though originating from different users, reflect convergent behavioral patterns that are likely to lead to the same recommendation target.

We treat such sequence pairs $(S_u, S_{u'})$ as positive examples for the Barlow Twins loss. This encourages the model to produce similar embeddings for distinct consumption paths that converge to the same intent, while still allowing diversity across user histories. Crucially, we avoid applying random augmentations (e.g., masking, cropping, dropout-based perturbations), as we find that they introduce noise and diminish performance in our multi-task, non-contrastive setup.

This augmentation strategy is natural for the next-item prediction task and emphasizes behavioral convergence as a signal for alignment, allowing the model to generalize across different interaction paths while preserving user-specific information.

We avoid synthetic augmentations (masking, cropping, dropout) proposed in~\cite{xie2021contrastivelearningsequentialrecommendation}, as they introduce stochastic noise, harm stability in our non-contrastive multi-task setup, and do not provide controllability in recommendation behavior.

\subsection{Redundancy-Reduction-Based Regularization}
\label{sec:bt}
Building on the behavioral alignment provided by our augmentation scheme, we now define a redundancy-reduction objective that enhances the quality of sequence embeddings by promoting feature diversity and invariance. This objective, based on the Barlow Twins (BT) framework, is integrated as an auxiliary loss in our multi-task training setup.

Let \(Z^A\) and \(Z^B\) denote the original and augmented batches of sequence embeddings produced by the backbone network. We assume both views are \(\ell_2\)-normalized and mean-centered across the batch. We apply \(\ell_2\) normalization instead of the learned projection head originally used in the BT framework, as it is parameter-free and empirically more stable. The cross-correlation matrix \(\mathcal{C}\in\mathbb{R}^{D\times D}\) is computed as
\begin{align}\label{eq:1}
C_{ij}
&= \frac{1}{B}
\sum_{b=1}^B
\frac{Z^A_{b,i}\;Z^B_{b,j}}
{\sqrt{\sum_{b'=1}^B (Z^A_{b',i})^2}\;\sqrt{\sum_{b'=1}^B (Z^B_{b',j})^2}},
\end{align}
where \(b\) indexes samples in a batch of size \(B\), and \(i,j\) index embedding dimensions. Each \(C_{ij}\in[-1,1]\), with 1 indicating perfect correlation and -1 indicating perfect anti-correlation.

The corresponding BT loss is composed of two terms:
\begin{align}\label{eq:2}
\mathcal{L}_{BT}
&= \sum_{i=1}^D (1 - C_{ii})^2
\;+\;\lambda\sum_{i=1}^D\sum_{\substack{j=1\\j\neq i}}^D C_{ij}^2,
\end{align}
where \(\lambda\) trades off invariance against decorrelation. Driving the diagonal elements of $C$ toward 1 enforces perturbation invariance, while pushing the off-diagonal elements toward 0 reduces redundancy. 

We now define the complete loss function used to train BT-SR, combining next-item prediction with redundancy reduction in a single multi-task objective:
\begin{equation}
\mathcal{L}_{\text{total}} = \mathcal{L}_{\text{pred}} + \alpha \mathcal{L}_{\text{BT}},
\label{eq:total_loss}
\end{equation}
where \( \mathcal{L}_{\text{pred}} \) is the standard next-item prediction loss and \( \mathcal{L}_{\text{BT}} \) is the Barlow Twins redundancy-reduction loss. The hyperparameter \( \alpha \) modulates the influence of self-supervised regularization during training.

This formulation enables the model to learn embeddings that are both robust to behavioral variation and structurally expressive. As shown in Section~\ref{sec:results}, tuning \( \alpha \) allows practitioners to steer recommendation behavior in a controllable and interpretable way.

\begin{table}[t]
\caption{NDCG@10 performance comparison of three SASRec variants (BCE loss, CE loss, Scalable CE loss) with and without the Barlow Twins regularization term. Results highlight consistent improvements from Barlow Twins across all scenarios.}
\label{table:utility_backbones}
    {\footnotesize
        \begin{tabular}{rccc|ccc}
        \toprule
         & BCE & FCE & SCE & BCE + BT & FCE + BT & SCE + BT \\
        \midrule
        ML1M & 0.0443 & 0.0494 & 0.0503 & 0.0490 & 0.0583 & 0.0562 \\
        YELP & 0.0146 & 0.0127 & 0.0137 & 0.0149 & 0.0150 & 0.0141 \\
        Gowalla & 0.0419 & 0.0364 & 0.0457 & 0.0434 & 0.0481 & 0.0483 \\
        Beauty & 0.0434 & 0.0524 & 0.0543 & 0.0440 & 0.0563 & 0.0549 \\
        Kindle Store & 0.0705 & 0.0700 & 0.0765 & 0.0733 & 0.0746 & 0.0796 \\
        \bottomrule
        \end{tabular}
    }
\end{table}

\begin{table*}[!t]
    \caption{Performance comparison across all datasets. Bold scores are the best on the dataset for the given metric, underlined scores are the second best.}
    {\footnotesize
    \begin{tabular}{rlcccccccl}
    \toprule
    Dataset & Metric & SasRec(BCE) & SasRec(CE) & SasRec(SCE) & CL4SRec & DuoRec & EC4SRec & \textbf{BT-SR}(Ours) & Improve \\
    \midrule
    \multirow{9}{*}{ML 1M}
 & hr@1    & 0.0160                   & \underline{0.0213}       & 0.0200                   & 0.0182                   & 0.0200                   & 0.0213                   & \textbf{0.0240* $ \pm $ 0.0048} & +12.7\% \\
 & hr@5    & 0.0493                   & \underline{0.0600}       & 0.0587                   & 0.0476                   & 0.0480                   & 0.0534                   & \textbf{0.0639* $ \pm $ 0.0024} & +6.5\% \\
 & hr@10   & 0.0893                   & 0.0847                   & \underline{0.0933}       & 0.0834                   & 0.0800                   & 0.0880                   & \textbf{0.0945* $ \pm $ 0.0080} & +1.3\% \\
  \arrayrulecolor{gray!30}\cmidrule(lr){2-10}\arrayrulecolor{black}

 & ndcg@5  & 0.0319                   & \underline{0.0404}       & 0.0393                   & 0.0342                   & 0.0353                   & 0.0364                   & \textbf{0.0436* $ \pm $ 0.0020} & +7.9\% \\ 
 
 & ndcg@10 & 0.0443                   & 0.0494                   & \underline{0.0503}       & 0.0464                   & 0.0456                   & 0.0487                   & \textbf{0.0534* $ \pm $ 0.0037} & +6.2\% \\
 & ndcg@50 & 0.0799                   & \underline{0.0818}       & 0.0800                   & 0.0712                   & 0.0723                   & 0.0748                   & \textbf{0.0833* $ \pm $ 0.0033} & +1.8\% \\
\arrayrulecolor{gray!30}\cmidrule(lr){2-10}\arrayrulecolor{black}
 & cov@1   & 0.0320                   & \underline{0.0887}       & \textbf{0.0961}          & 0.0335                   & 0.0361                   & 0.0387                   & 0.0816 $ \pm $ 0.0035          & -15.1\% \\
 & cov@5   & 0.1473                   & \underline{0.2265}       & \textbf{0.2568}          & 0.0969                   & 0.0936                   & 0.1002                   & 0.2223 $ \pm $ 0.0083          & -13.4\% \\
 & cov@10  & 0.2434                   & 0.3185                   & \textbf{0.3645}          & 0.1329                   & 0.1459                   & 0.1591                   & \underline{0.3189 $ \pm $ 0.0123}& -12.5\% \\
    \midrule
    \multirow{9}{*}{YELP}
     & hr@1   & \underline{0.0045} & 0.0043 & 0.0040 & 0.0028 & 0.0027 & 0.0028 & \textbf{0.0049* $\pm$ 0.0005}  & +8.9\% \\
     & hr@5   & \underline{0.0162} & 0.0137 & 0.0152 & 0.0087 & 0.0094 & 0.0090 & \textbf{0.0183* $\pm$ 0.0008}  & +13.0\% \\
     & hr@10  & \textbf{0.0292} & 0.0257 & 0.0277 & 0.0192 & 0.0188 & 0.0199 & \underline{0.0288 $\pm$ 0.0010}  & –1.4\% \\ \arrayrulecolor{gray!30}\cmidrule(lr){2-10}\arrayrulecolor{black}
     & ndcg@5 & \underline{0.0104} & 0.0088 & 0.0097 & 0.0060 & 0.0061 & 0.0062 & \textbf{0.0117* $\pm$ 0.0006}  & +12.5\% \\
     & ndcg@10& \underline{0.0146} & 0.0127 & 0.0137 & 0.0085 & 0.0091 & 0.0092 & \textbf{0.0150* $\pm$ 0.0006}  & +2.7\%  \\
     & ndcg@50& \textbf{0.0280} & 0.0263 & 0.0258 & 0.0178 & 0.0172 & 0.0184 & \underline{0.0267 $\pm$ 0.0007}  & –4.6\%  \\ \arrayrulecolor{gray!30}\cmidrule(lr){2-10}\arrayrulecolor{black}
     & cov@1  & 0.0103 & \textbf{0.0233} & 0.0135 & 0.0057 & 0.0059 & 0.0059 & \underline{0.0218 $\pm$ 0.0034}  & –6.4\%  \\
     & cov@5  & 0.0333 & \underline{0.0621} & 0.0413 & 0.0170 & 0.0161 & 0.0171 & \textbf{0.0636* $\pm$ 0.0089}  & +2.4\%  \\
     & cov@10 & 0.0541 & \underline{0.0911} & 0.0661 & 0.0276 & 0.0255 & 0.0275 & \textbf{0.0998* $\pm$ 0.0128}  & +9.5\% \\
    \midrule
    \multirow{9}{*}{Gowalla}
     & hr@1   & 0.0173 & 0.0143 & \underline{0.0196} & 0.0157 & 0.0144 & 0.0142 & \textbf{0.0205* $\pm$ 0.0010} & +4.6\%  \\
     & hr@5   & 0.0506 & 0.0431 & \underline{0.0548} & 0.0433 & 0.0473 & 0.0440 & \textbf{0.0592* $\pm$ 0.0010}  & +8.0\%  \\
     & hr@10  & 0.0756 & 0.0660 & \underline{0.0811} & 0.0694 & 0.0712 & 0.0757 & \textbf{0.0851* $\pm$ 0.0020}  & +4.9\%  \\ \arrayrulecolor{gray!30}\cmidrule(lr){2-10}\arrayrulecolor{black}
     & ndcg@5 & 0.0339 & 0.0289 & \underline{0.0372} & 0.0318 & 0.0310 & 0.0331 & \textbf{0.0400* $\pm$ 0.0009}  & +7.5\%  \\
     & ndcg@10& 0.0419 & 0.0364 & \underline{0.0457} & 0.0411 & 0.0387 & 0.0397 & \textbf{0.0483* $\pm$ 0.0013}  & +5.7\%  \\
     & ndcg@50& 0.0613 & 0.0528 & \underline{0.0640} & 0.0535 & 0.0587 & 0.0549 & \textbf{0.0667* $\pm$ 0.0015}  & +4.2\%  \\ \arrayrulecolor{gray!30}\cmidrule(lr){2-10}\arrayrulecolor{black}
     & cov@1  & 0.0230 & 0.0228 & 0.0214 & 0.0252 & 0.0255 & \underline{0.0270} & \textbf{0.0321* $\pm$ 0.0039}  & +25.9\% \\
     & cov@5  & 0.0892 & 0.0725 & 0.0921 & 0.0950 & 0.0939 & \underline{0.0998} & \textbf{0.1280* $\pm$ 0.0140}  & +36.3\% \\
     & cov@10 & 0.1557 & 0.1153 & 0.1625 & 0.1538 & 0.1595 & \underline{0.1650} & \textbf{0.2187* $\pm$ 0.0237}  & +34.6\% \\
    \midrule
    \multirow{9}{*}{Beauty}
 & hr@1    & 0.0179                   & 0.0269                   & \underline{0.0305}       & 0.0292                   & 0.0305                   & 0.0303                   & \textbf{0.0326* $ \pm $ 0.0013} & +6.9\% \\
 & hr@5    & 0.0538                   & \underline{0.0591}       & 0.0582                   & 0.0555                   & 0.0573                   & 0.0574                   & \textbf{0.0606* $ \pm $ 0.0025} & +2.5\% \\
 & hr@10   & 0.0789                   & 0.0896                   & \underline{0.0905}       & 0.0830                   & 0.0860                   & 0.0815                   & \textbf{0.0937* $ \pm $ 0.0019} & +3.5\% \\
\arrayrulecolor{gray!30}\cmidrule(lr){2-10}\arrayrulecolor{black}
 & ndcg@5  & 0.0355                   & 0.0425                   & 0.0442                   & 0.0412                   & 0.0436                   & \underline{0.0455}       & \textbf{0.0463* $ \pm $ 0.0002} & +1.8\% \\
 & ndcg@10 & 0.0434                   & 0.0524                   & \underline{0.0543}       & 0.0530                   & 0.0527                   & 0.0537                   & \textbf{0.0569* $ \pm $ 0.0001} & +4.8\% \\
 & ndcg@50 & 0.0659                   & 0.0777                   & 0.0794                   & \textbf{0.0833}          & 0.0789                   & 0.0794                   & \underline{0.0813 $ \pm $ 0.0035}& -2.4\% \\
\arrayrulecolor{gray!30}\cmidrule(lr){2-10}\arrayrulecolor{black}
 & cov@1   & \underline{0.0620}       & 0.0546                   & 0.0499                   & 0.0603                   & 0.0607                   & \textbf{0.0679}          & 0.0491 $ \pm $ 0.0078          & -20.7\% \\
 & cov@5   & \underline{0.2200}       & 0.1782                   & 0.1760                   & \textbf{0.2212}          & 0.2100                   & 0.2103                   & 0.1532 $ \pm $ 0.0357          & -29.2\% \\
 & cov@10  & \textbf{0.3448}          & 0.2752                   & 0.2766                   & \underline{0.3362}       & 0.3234                   & 0.3236                   & 0.3234 $ \pm $ 0.0612          & -30.5\% \\
    \midrule
    \multirow{9}{*}{Kindle Store}
 & hr@1    & 0.0451                   & 0.0471                   & \underline{0.0533}       & 0.0469                   & 0.0471                   & 0.0505                   & \textbf{0.0564* $ \pm $ 0.0007} & +5.8\% \\
 & hr@5    & 0.0832                   & 0.0813                   & \underline{0.0899}       & 0.0813                   & 0.0762                   & 0.0785                   & \textbf{0.0903* $ \pm $ 0.0018} & +0.4\% \\
 & hr@10   & 0.0990                   & 0.0965                   & \underline{0.1022}       & 0.0910                   & 0.0897                   & 0.0955                   & \textbf{0.1058* $ \pm $ 0.0012} & +3.5\% \\ \arrayrulecolor{gray!30}\cmidrule(lr){2-10}\arrayrulecolor{black}
 & ndcg@5  & 0.0655                   & 0.0652                   & \underline{0.0726}       & 0.0635                   & 0.0625                   & 0.0652                   & \textbf{0.0747* $ \pm $ 0.0009} & +2.9\% \\
 & ndcg@10 & 0.0705                   & 0.0700                   & \underline{0.0765}       & 0.0662                   & 0.0668                   & 0.0672                   & \textbf{0.0796* $ \pm $ 0.0005} & +4.1\% \\
 & ndcg@50 & 0.0786                   & 0.0803                   & \underline{0.0852}       & 0.0731                   & 0.0771                   & 0.0736                   & \textbf{0.0883* $ \pm $ 0.0007} & +3.6\% \\ \arrayrulecolor{gray!30}\cmidrule(lr){2-10}\arrayrulecolor{black}
 & cov@1   & 0.0396                   & 0.0337                   & \underline{0.0409}       & 0.0363                   & 0.0356                   & 0.0369                   & \textbf{0.0440* $ \pm $ 0.0005} & +7.6\% \\
 & cov@5   & 0.1278                   & 0.1076                   & \underline{0.1492}       & 0.1077                   & 0.1123                   & 0.1082                   & \textbf{0.1665* $ \pm $ 0.0019} & +11.6\% \\
 & cov@10  & 0.1920                   & 0.1665                   & \underline{0.2346}       & 0.1598                   & 0.1719                   & 0.1703                   & \textbf{0.2673* $ \pm $ 0.0021} & +13.9\% \\
    \bottomrule
\end{tabular}

    }
    \label{table:utility_metrics}
\end{table*}

\section{Experimental Setup}

\textbf{Datasets.} 
We follow the experimental protocol proposed in~\cite{Mezentsev_2024} for benchmarking backbone architectures. Experiments are conducted on five public datasets: Behance~\cite{behance}, Kindle Store~\cite{ni-etal-2019-justifying}, Yelp~\cite{asghar2016yelp}, Gowalla~\cite{gowalla}, and MovieLens-1M~\cite{movielens2015}. 
Users and items with fewer than five interactions are filtered out. 
To prevent temporal leakage, we adopt a timestamp-based split strategy~\cite{Frolov2022TensorBasedSL}. 
A global timestamp at the 0.95 quantile of all interactions defines the boundary between training and test data: interactions before this point form the training set, and users with subsequent interactions (excluded from training) constitute the test pool. 
For each test user, we apply a standard leave-one-out protocol, using the latest interaction for testing and the second-to-last interaction for validation. 
All experiments are conducted on a single NVIDIA A100 GPU. 
The code and preprocessing scripts are publicly available\footnote{\url{https://github.com/RAZVOR/barlow_twins_sasrec}}.

\textbf{Metrics.} 
Following best practices~\cite{ca2020, dallmann2021, krichene2020}, we evaluate models using unsampled top-$K$ ranking metrics computed over the full item catalog. 
We report Normalized Discounted Cumulative Gain (ndcg@K) and Hit Rate (hr@K) for \mbox{$K = 1, 5, 10, 50$}, averaged across all test users. 
To assess recommendation diversity, we report item coverage (cov@K), defined as the fraction of unique items appearing in top-$K$ recommendations across all users. 
In addition to ranking and coverage metrics, we analyze the structure of learned user embeddings using the \textit{effective rank}~\cite{7098875}. 
Given the singular values $\sigma_1, \sigma_2, \dots, \sigma_D$ of the user-embedding matrix (obtained via singular value decomposition), 
we normalize them as $p_i = \sigma_i / \sum_{j=1}^{D} \sigma_j$ and define the effective rank as the exponential of the Shannon entropy:
\begin{equation}
r_{\mathrm{eff}} = \exp\!\Big(-\sum_{i=1}^{D} p_i \log p_i\Big).
\label{eq:eff_rank}
\end{equation}
A higher $r_{\mathrm{eff}}$ indicates a flatter singular-value spectrum and therefore a more diverse, less redundant embedding space.

\textbf{Hyperparameters.} 
All models are implemented in PyTorch and optimized with Adam using a learning rate of 0.001. 
The maximum sequence length is fixed to 50, truncating longer histories to the most recent interactions. 
L2 weight decay is applied for regularization, with the coefficient tuned on the validation set (typically \(10^{-4}\)–\(10^{-5}\)).
We perform a grid search over the Barlow Twins hyperparameters $\alpha$ and $\lambda$, selecting values from \(\{0.05, 0.10, \dots, 0.50\}\) based on validation performance. 
The auxiliary loss \( \mathcal{L}_{BT} \) is applied to different SASRec variants trained with \( \mathcal{L}_{\text{CE}} \), \( \mathcal{L}_{\text{Scalable CE}} \)~\cite{Mezentsev_2024}, and \( \mathcal{L}_{\text{BCE}} \)~\cite{dross2023}. 
We report the best-performing model for each dataset. 
To ensure statistical significance, we report the mean and standard deviation over five independent runs. 
Significance is verified using paired t-tests ($p < 0.05$) against the second-best baseline, with statistically significant improvements marked by an asterisk (*). 
The low observed variance demonstrates the stability of our method.

\subsection{Baselines}
\label{subsec:baselines}

As baselines, we adopt SASRec variants trained with: (1) binary cross-entropy (BCE)~\cite{DBLP:journals/corr/abs-1808-09781}, (2) standard cross-entropy (CE), and (3) scalable cross-entropy (SCE)~\cite{Mezentsev_2024}. 
For contrastive-learning baselines, we include CL4SRec~\cite{xie2021contrastivelearningsequentialrecommendation}, DuoRec~\cite{DBLP:journals/corr/abs-2110-05730}, and EC4Rec, an explanation-guided contrastive framework that leverages training gradients to identify positive and negative items~\cite{wang2022explanationguidedcontrastivelearning}. 

While recent LLM- and LoRA-based recommenders perform well in text-rich or cold-start settings, they underperform in ID-based domains and incur substantially higher inference costs. 
Moreover, prior studies~\cite{10.1145/3726302.3730178, DBLP:journals/corr/abs-2503-05493} show that large language models often memorize public recommendation datasets, raising concerns of target leakage rather than genuine generalization. 
For these reasons, we exclude LLM-based baselines from our evaluation.

\section{Results}
\label{sec:results}
% \subsection{Overall Performance}

\begin{figure}[!t]
    \centering

    % ─── first row ─────────────────────────────────────────────
    \includegraphics[width=0.5\linewidth]{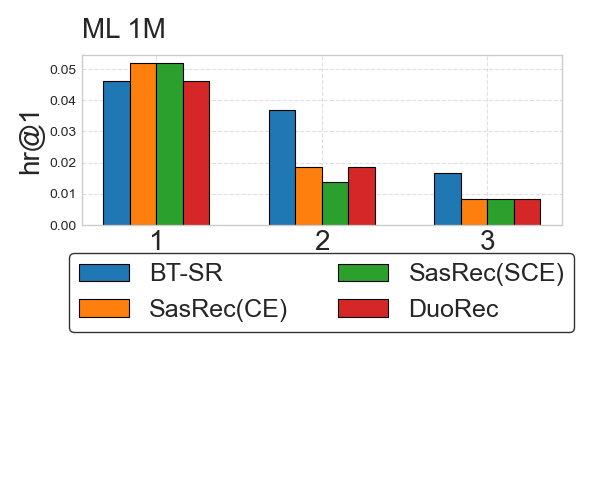}
    \includegraphics[width=0.49\linewidth]{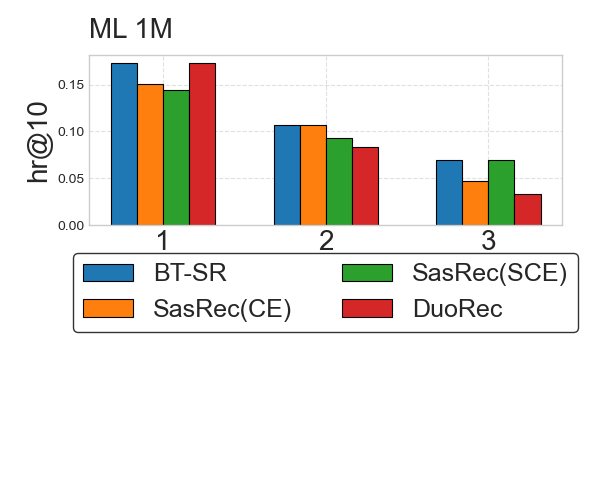}
    \\[-30pt]
    
    \includegraphics[width=0.5\linewidth]{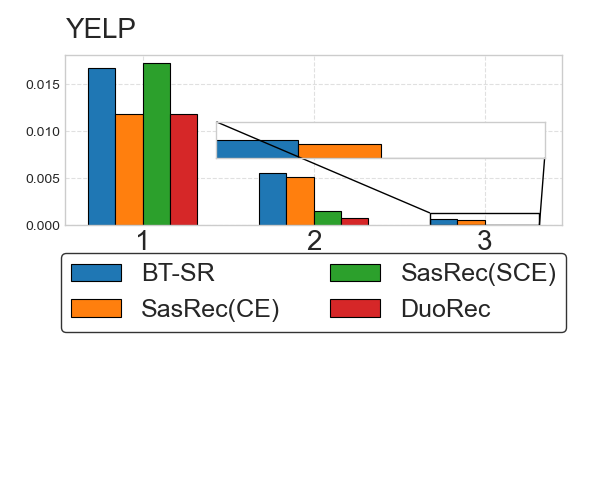}
    \includegraphics[width=0.49\linewidth]{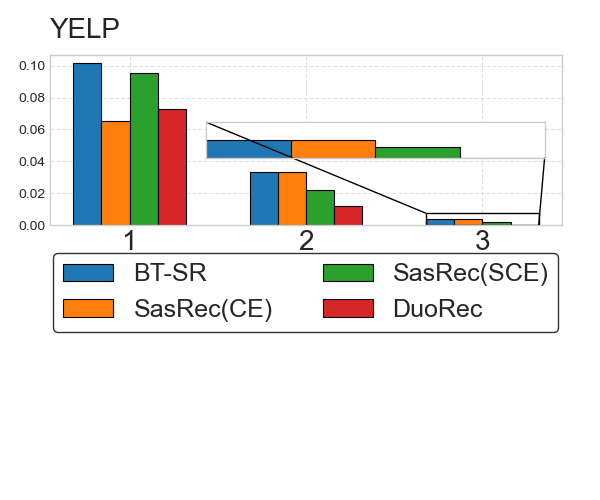}
    \\[-30pt]
    
    \includegraphics[width=0.5\linewidth]{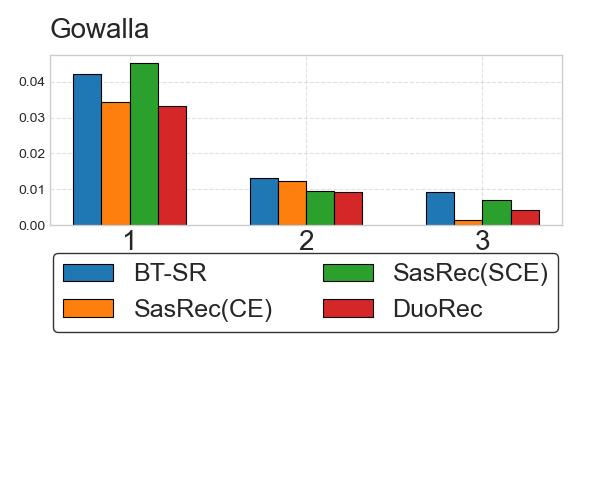}
    \includegraphics[width=0.49\linewidth]{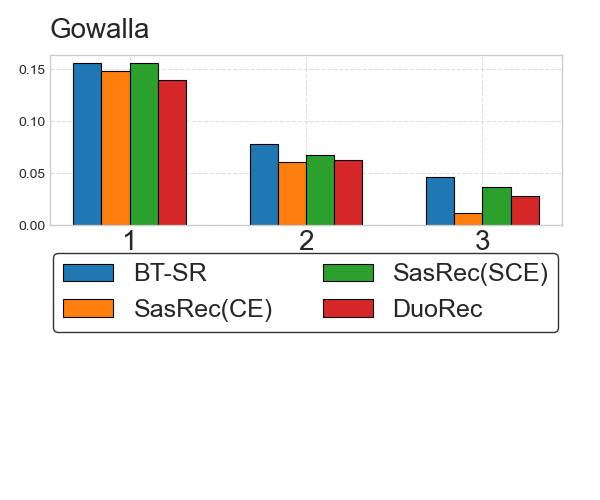}
    \\[-32pt]
    
    \caption{HR@1 (left) and HR@10 (right) metrics for the three item-popularity buckets across three datasets.}
    \label{figure:pop_buckets}
    %\\[-20pt]
\end{figure}

We first evaluate the impact of adding the Barlow Twins term to three SasRec objectives (\(\mathcal{L}_{BCE}\), \(\mathcal{L}_{CE}\), \(\mathcal{L}_{SCE}\)). As shown in Table~\ref{table:utility_backbones}, the Barlow Twins term consistently improves performance across all base losses in the multi-task setup, demonstrating the universality of our approach.  

Next, we select the best-performing objective for each dataset. \(\mathcal{L}_{SCE}\) achieves the strongest results on Gowalla and Kindle Store, while \(\mathcal{L}_{CE}\) performs better on the remaining datasets. Using these selected models, we compare against strong baselines on the holdout test set. Table~\ref{table:utility_metrics} reports our best variant (BT-SR), which consistently outperforms all baselines in both utility and coverage across most datasets. To further investigate the role of the Barlow Twins loss in representation learning, we provide a detailed analysis in the following section.

% \section{Results analysis}
% \vspace{5em}
\subsection{Analysis}

\subsubsection{What Drives BT-SR’s Outperformance?} \;

Although BT-SR achieves only modest coverage gains on MovieLens and Amazon Beauty—remaining below the second-best baseline—it delivers substantially higher coverage on the other three datasets. To unpack this behavior, we split each item corpus into three popularity-based buckets (each containing roughly one-third of total interactions), ordered from most to least popular. Figure~\ref{figure:pop_buckets} shows HR@1 and HR@10 for each bucket.

Surprisingly, BT-SR underperforms slightly at HR@1 in the top-popularity bucket, yet it markedly outperforms all baselines on the mid- and low-popularity buckets—evidence of its enhanced personalization. Moreover, BT-SR also leads at HR@10 even for the most popular items, indicating that \emph{it can elevate niche items into top-rank positions without sacrificing performance on blockbusters}. This pattern highlights how the Barlow Twins loss both prevents representation collapse and \emph{mitigates popularity bias}.

Further, in Section 5.4 we demonstrate that BT-SR can be further tuned for industrial applications via an optional hyperparameter that explicitly balances HR@1 against HR@10—allowing practitioners to prioritize the metric that best suits their use case.

\begin{figure}[!t]
    \centering

    \includegraphics[width=0.5\linewidth]{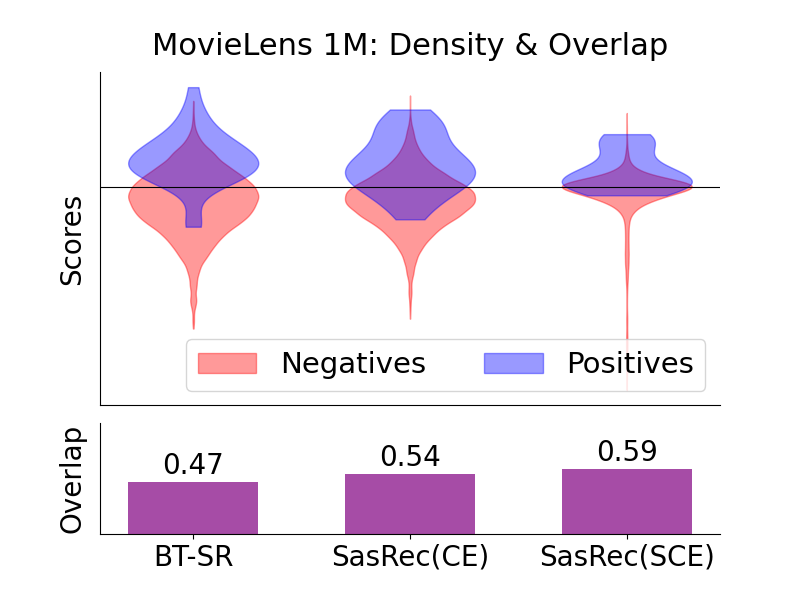}
    \includegraphics[width=0.49\linewidth]{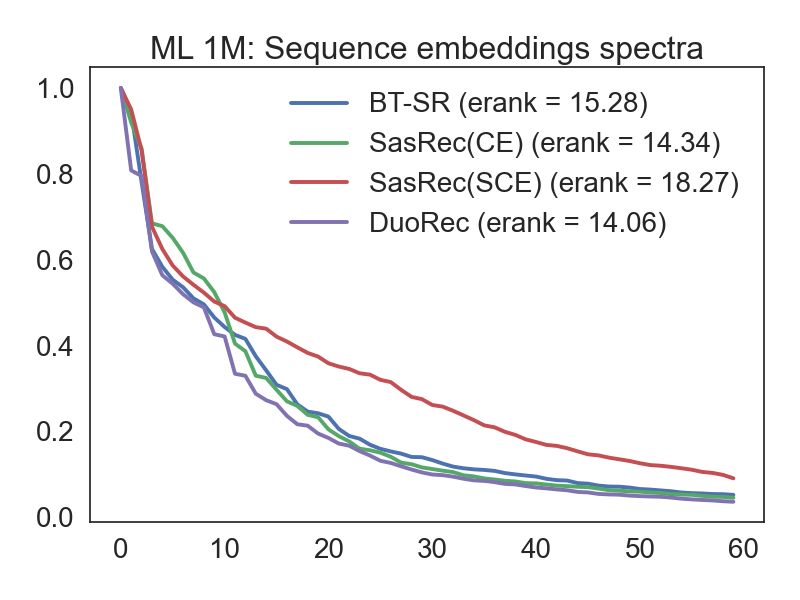}
    
    \includegraphics[width=0.49\linewidth]{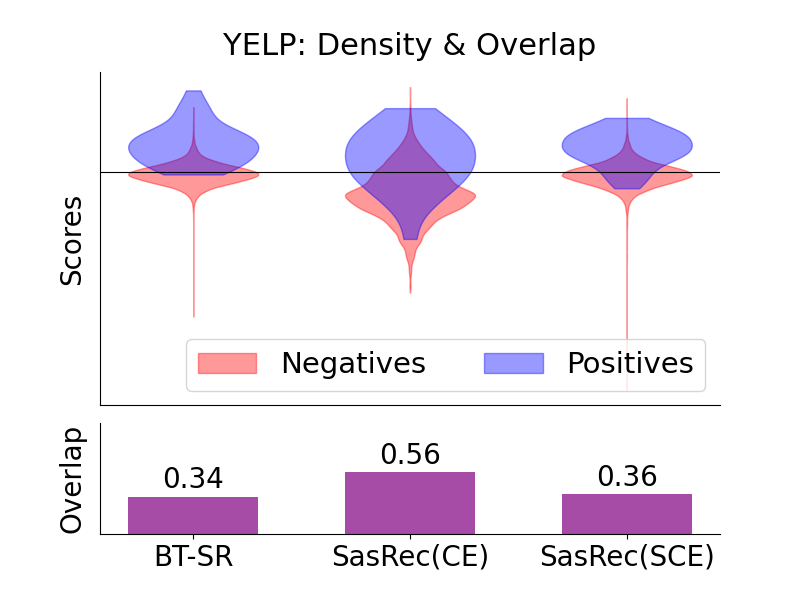}
    \includegraphics[width=0.49\linewidth]{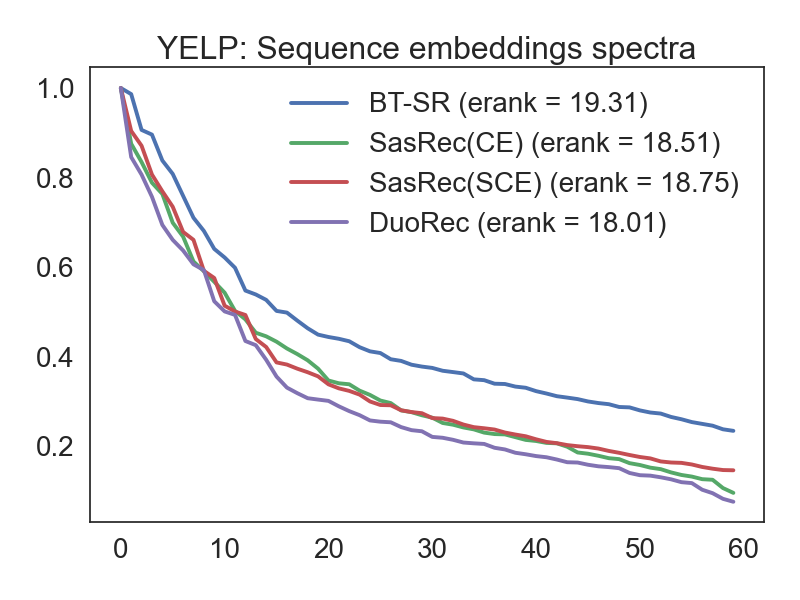}
    
    \includegraphics[width=0.5\linewidth]{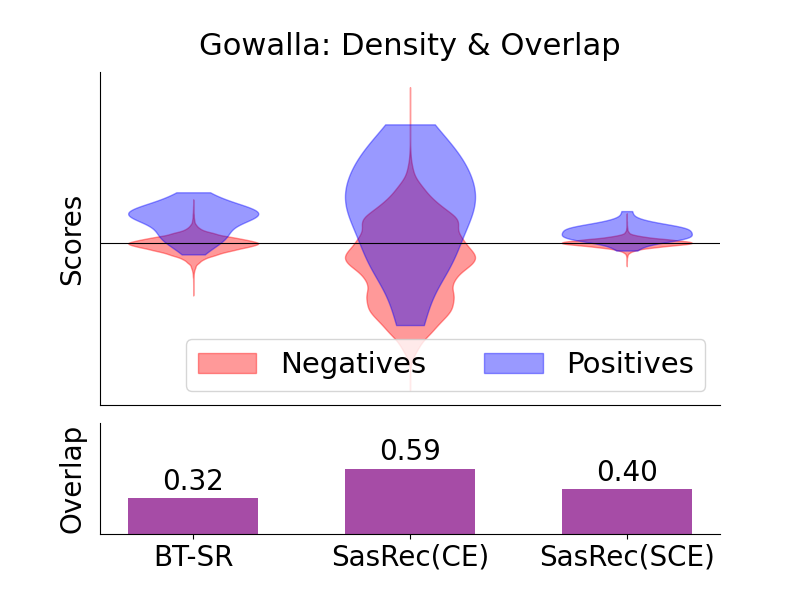}
    \includegraphics[width=0.49\linewidth]{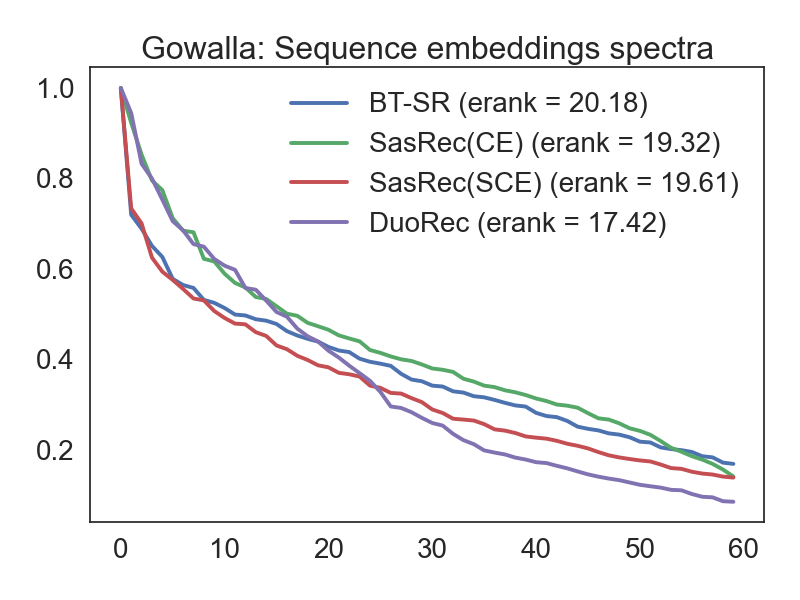}
    
    \caption{(left) Comparison of score density distributions for positive and negative candidate pairs across three datasets. We quantify model’s confidence in distinguishing relevant candidates by the histogram overlap factor (\emph{Overlap})—lower values indicate better separation. (right) Singular value spectra of the sequence embeddings for each dataset, annotated with their computed effective ranks (in legend), illustrate the effective dimensionality of the learned representations. }
    \label{figure:scores_dens_fig}
    %\\[-10pt]
\end{figure}

\subsubsection{Can non-contrastive learning produce higher-quality user embedding spaces than contrastive learning in recommendation?} \;

Our method outperforms contrastive learning (CL) baselines, not only demonstrating greater effectiveness but also suggesting that it learns higher-quality user embeddings.

We compared the score distributions for positive (ground-truth) and negative items. Figure~\ref{figure:scores_dens_fig} shows that BT-SR consistently assigns higher scores to positive items than competing models across three datasets. This confirms that our approach sharpens user representations and more clearly distinguishes true preferences from noise.

To analyze the embedding geometry, we performed singular value decomposition (SVD) on each set of user embeddings and plotted their singular-value spectra (Figure~\ref{figure:scores_dens_fig}). We then computed the effective rank (Eq.~\ref{eq:eff_rank}) to quantify the flatness of the singular-value spectrum. On Yelp and Gowalla, BT-SR produces a flatter spectrum, whereas on MovieLens the spectrum is steeper.

\begin{figure}[!t]
    \centering

    % ─── first row ─────────────────────────────────────────────
    \includegraphics[width=0.42\linewidth]{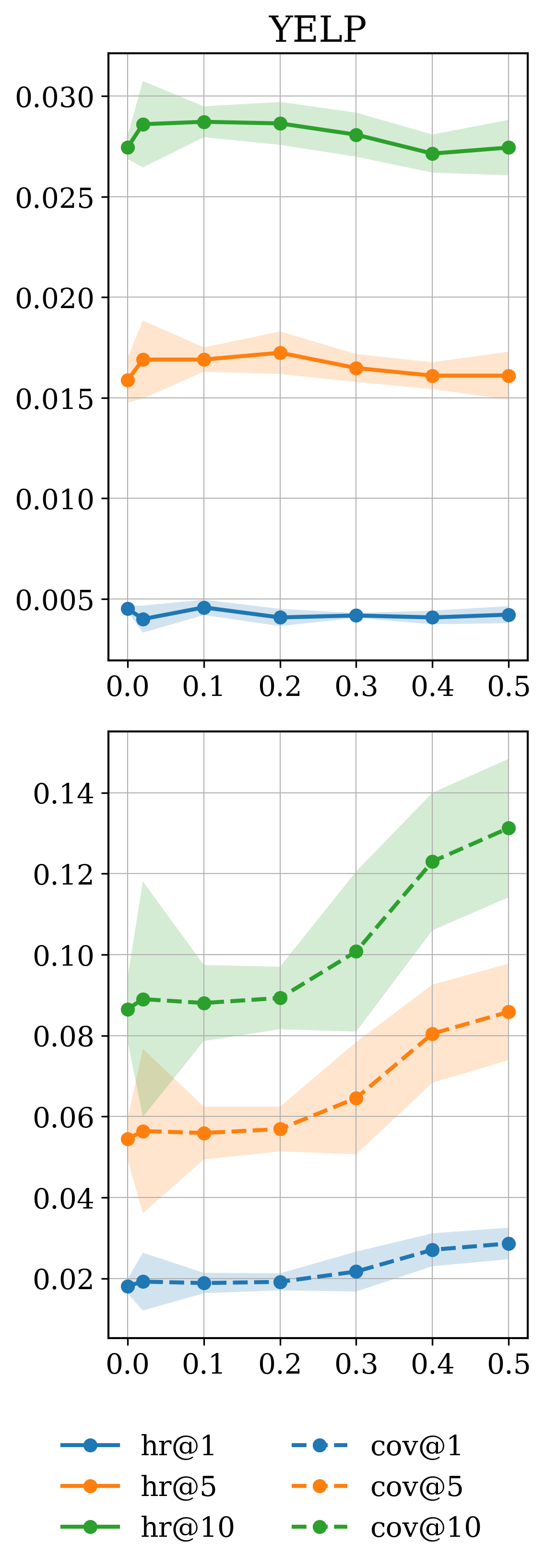}
    \includegraphics[width=0.49\linewidth]{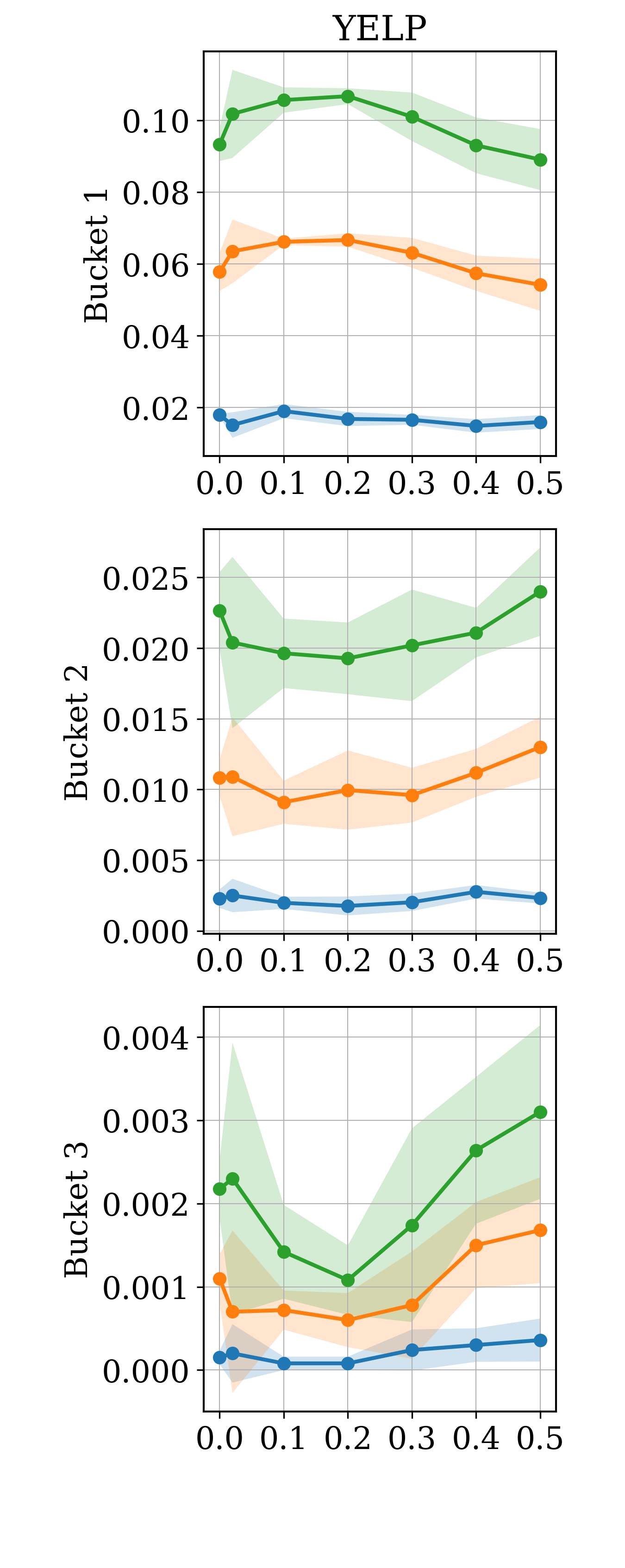}
    
    \caption{Parameter sensitivity wrt $\alpha$ for YELP dataset. Left: common metrics, Right: item-popularity bucket metrics. Picture demonstrates controllability of recommendations via hyperparameter. }
    \label{figure:yelp_abol}
\end{figure}

\begin{figure}[!t]
    \centering

    % ─── first row ─────────────────────────────────────────────
    % \begin{minipage}[t]{0.49\linewidth}
    %   \includegraphics[width=\linewidth]{pictures/ablation/ML1M_lambda_1_cov.png}
    % \end{minipage}
    % \hfill
    % \begin{minipage}[t]{0.49\linewidth}
    %   \includegraphics[width=\linewidth]{pictures/ablation/ML1M_lambda_1_buckets_hr.png}
    % \end{minipage}
    \includegraphics[width=0.42\linewidth]{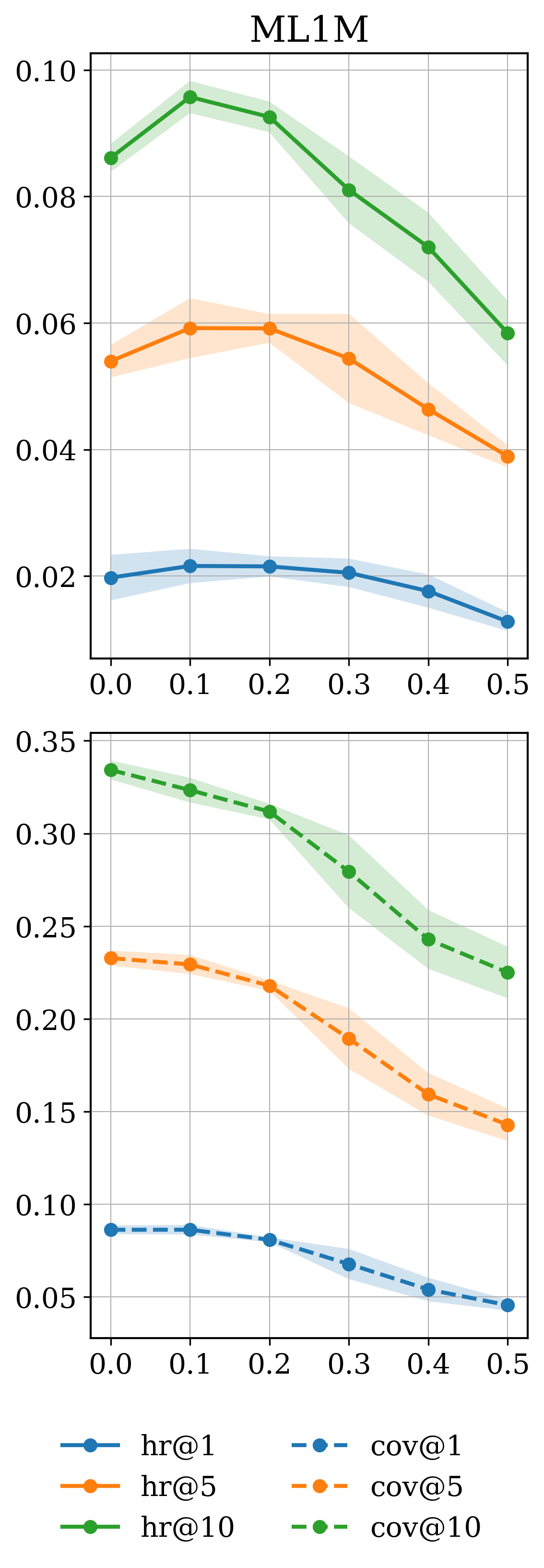}
    \includegraphics[width=0.49\linewidth]{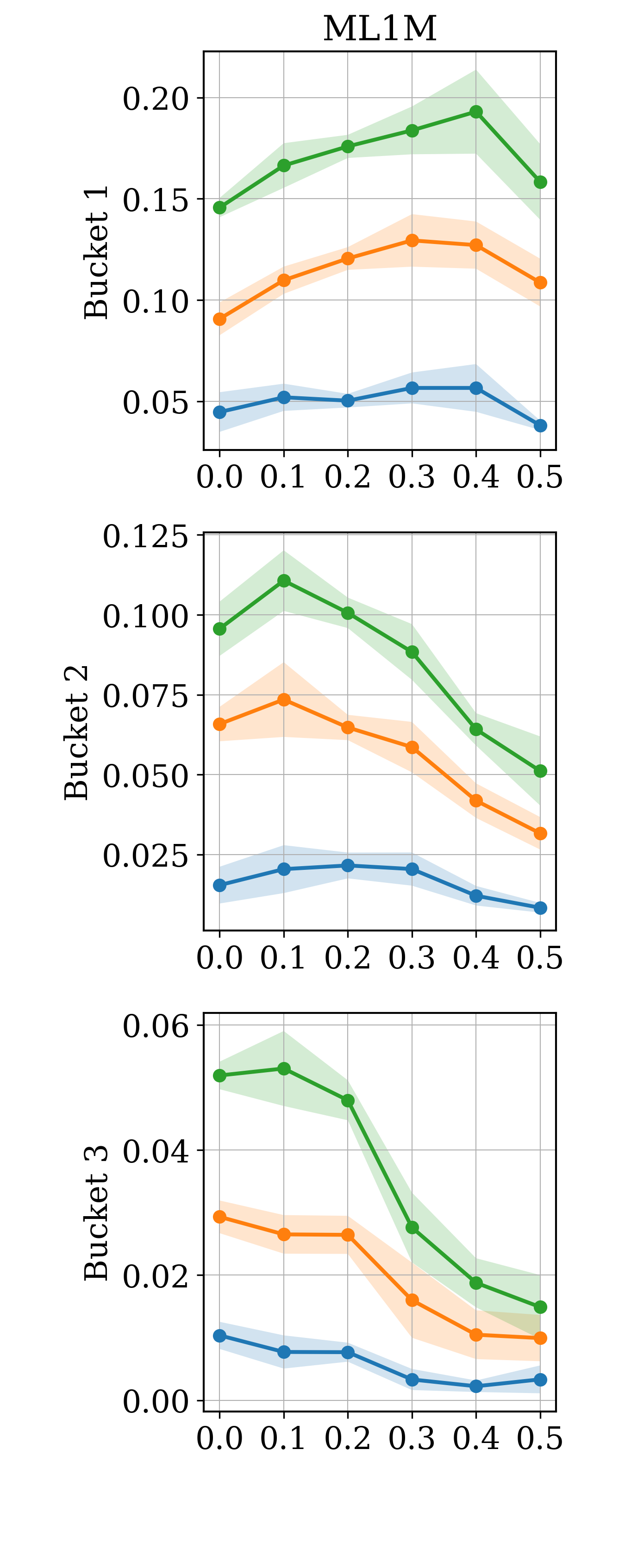}
    
    \caption{Parameter sensitivity wrt $\alpha$ for MovieLens dataset. Left: common metrics, Right: item-popularity bucket metrics. Picture demonstrates controllability of recommendations via hyperparameter. }
    \label{figure:ml_abol}
\end{figure}

\subsection{Ablation Study}

To analyze the balance between the primary recommendation loss and our Barlow–Twins regularizer, we performed a hyperparameter sweep over
\(
\alpha \;\in\;\{0.0, 0.1, 0.2, 0.3, 0.4, 0.5\},
\)
where \(\alpha=0\) effectively removes the BT term in Equation~\ref{eq:total_loss}. We also evaluated sensitivity to the decorrelation weight \(\lambda\) in Equation~(11). To isolate each effect, we first fixed \(\lambda\) at its optimal value and varied \(\alpha\), and then held \(\alpha\) constant while sweeping \(\lambda\). Figures~\ref{figure:yelp_abol} and~\ref{figure:ml_abol} plot (i) hit‐rate metrics (\(\mathrm{hr}@1\), \(\mathrm{hr}@5\), \(\mathrm{hr}@10\)) and (ii) coverage@\(K\) as functions of \(\alpha\). Additional ablation studies are available in Appendix A.

When \(\alpha \le 0.1\), the model favors a short, high‐precision recommendation list: \(\mathrm{hr}@1\) and \(\mathrm{hr}@5\) peak, while coverage remains low. As \(\alpha\) increases, precision at top‐1 drops by only 2–3 pp, but \(\mathrm{hr}@10\) improves and coverage@10 nearly doubles. In effect, \(\alpha\) acts as a dial between a \emph{short‐list/high‐precision} mode and a \emph{long‐list/high‐diversity} mode.

To further dissect this trade‐off, we studied metrics for 3 item-popularity buckets. Figures~\ref{figure:ml_abol} and~\ref{figure:yelp_abol} show \(\mathrm{hr}@K\) and coverage@\(K\) within each bucket as \(\alpha\) varies. At low \(\alpha\), almost all hits come from Bucket 1 and Bucket 3 performance is near zero. As \(\alpha\) exceeds 0.2, performance in Buckets 2 and 3 rises sharply (e.g., coverage@10 in Bucket 3 grows from ~0 to ~0.015), while Bucket 1 performance declines only modestly.

We observe the same qualitative behavior on Yelp, Gowalla, and Amazon Beauty: a low‐\(\alpha\) regime tuned for head‐item precision versus a high‐\(\alpha\) regime that enhances long‐tail coverage. MovieLens (ML-1M), however, shows weaker bucket separation: decreasing \(\alpha\) improves head‐bucket metrics but degrades both mid‐ and tail‐bucket performance in tandem. We attribute this to the extreme interaction skew in MovieLens, which makes our uniform bucket split conflate heterogeneous user subgroups and obscure the head‐vs‐tail trade‐off.

\begin{figure}[!t]
    \centering

    % ─── first row ─────────────────────────────────────────────
    \vspace{-1em}
    \includegraphics[width=0.5\linewidth]{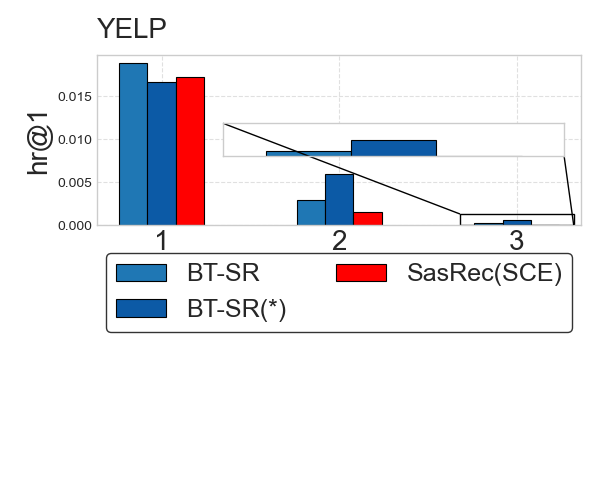}
    \includegraphics[width=0.49\linewidth]{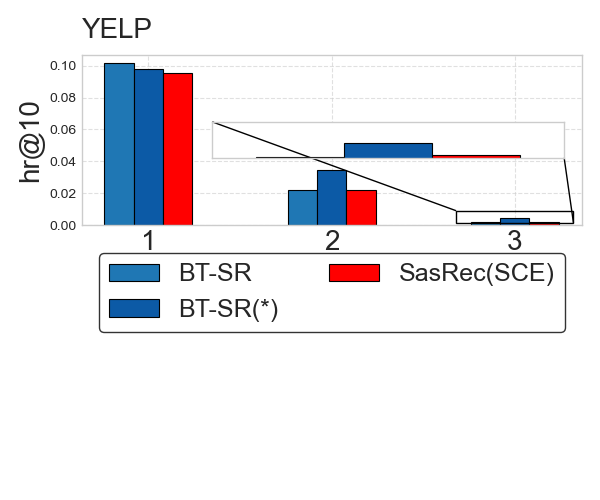}
    
    \caption{Performance comparison (HR@1 and HR@10) for the BT-SR method on the Yelp dataset under two regimes: $\alpha=0.1$ favors popular items (high HR@1), while $\alpha=0.4$ promotes diverse, less popular items early in the list, boosting personalization and maintaining high overall accuracy (HR@10).}
    \label{figure:pop_buckets_modes}
    %\\[-20pt]
\end{figure}

We further demonstrate the practical aspect of this effect on Figure~\ref{figure:pop_buckets_modes}. We provide two setups of our BT-SR method with two different values of $\alpha$ on the Yelp dataset. Both setups allow outperforming the baseline in terms of integral recommendations accuracy yet they yield completely different internal structure of recommendations with respect to item popularity. As demonstrated in the figure, one can select between two regimes. In one regime, corresponding to the lower value of $\alpha$, the recommendations are steered towards more generic user interests, which is indicated by a higher performance in the first bucket in terms of \(\mathrm{hr}@1\) metric. In contrast, the second regime with higher $\alpha$ compensates for lower scores in the first bucket by a better performance in the second and third buckets, thus promoting less popular yet relevant recommendations at the beginning of the recommendation list. It helps increasing the diversity of recommendations without compromising the overall accuracy and boosts personalization.

In summary, our framework not only improves overall recommendation quality but also exposes a single, intuitive hyperparameter \(\alpha\) that practitioners can tune to balance short‐list precision against long‐tail diversity.  
In practice, we find that moderate values of $\alpha$ (e.g., 0.2–0.4) offer the best balance across datasets.

\textbf{Impact of Augmentation Strategy.}
We next assess the impact of our behaviorally aligned augmentation strategy. 
Following prior contrastive learning approaches such as CL4SRec~\cite{DBLP:journals/corr/abs-2010-14395}, 
we experiment with perturbation-based augmentations (e.g., random masking, cropping, and reordering). 
However, we observe that these noise-based augmentations lead to performance degradation in our non-contrastive, multi-task setting. 
We further evaluate item-based augmentations on the YELP and Kindle Store datasets and find a performance drop of at least 6\%. 
These findings support our choice of goal-preserving augmentations derived from user intent, as described in Section~\ref{subsec:augs}.

\begin{table}[!t]
    \centering
    \begin{tabular}{c|c c c}
        \toprule
          & SASRec(BCE) & BT-SR( $ \alpha = 0.1 $ ) & BT-SR( $ \alpha = 0.4 $ ) \\ 
         \toprule
         \textbf{ndcg@20} & 0.0176 & 0.0180 & 0.0182 \\
         \toprule
    \end{tabular}
    \caption{Integral metric for Figure 7}
    \label{tab:placeholder}
\end{table}

\section{Conclusion}

We introduce BT-SR, a novel integration of the Barlow Twins redundancy-reduction objective into Transformer-based sequential recommenders. BT-SR outperforms six strong baselines, mitigates popularity bias, and produces sharper, more discriminative user embeddings. Moreover, our method allows adapting to different trade-offs between accuracy and diversity of recommendations, therefore better suiting various user engagement scenarios used in real applications. These results demonstrate that non-contrastive self-supervision offers a compact, effective pathway toward fairer, high-performing recommendation models.
% These findings highlight non-contrastive learning as a powerful, efficient, and more controllable alternative in sequential recommendation.

% \newpage
\printbibliography

\newpage
\appendix
\section{Ablation Study}

\begin{figure*}[h]
    \centering
    \includegraphics[width=0.99\linewidth]{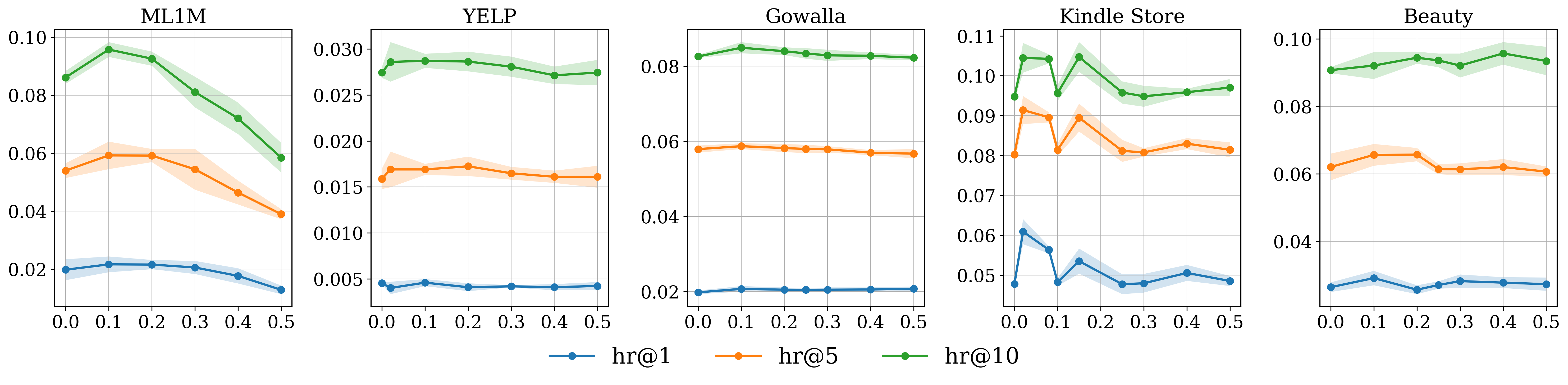}
    \includegraphics[width=0.99\linewidth]{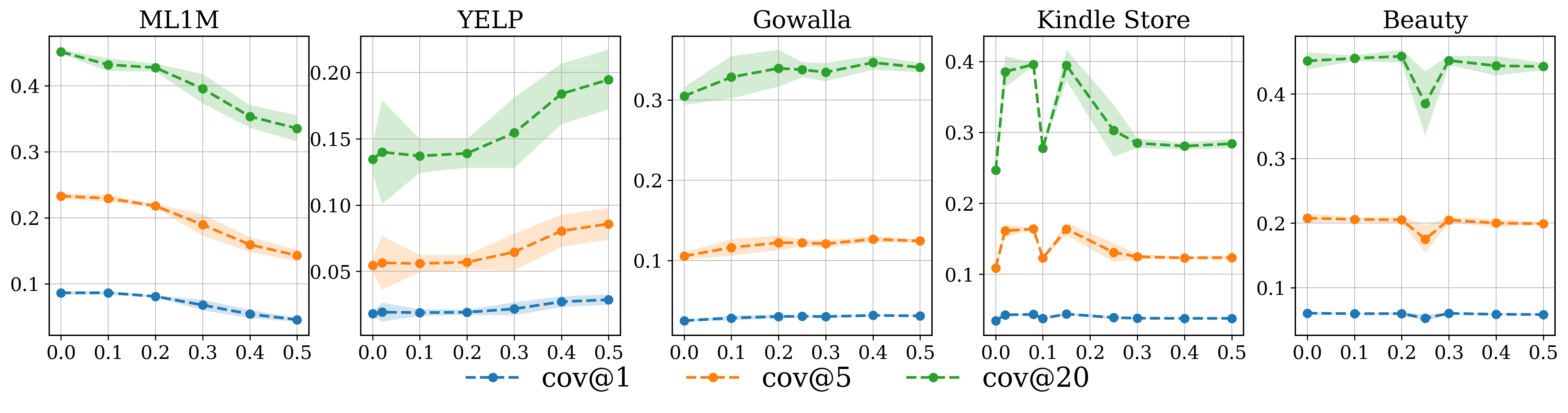}
    \caption{Sensitivity Analysis w.r.t. $ \alpha $ of hr@K and cov@K metrics }
    \label{figure:ablation_main}
\end{figure*}

Figure \ref{figure:ablation_main} presents the complete results of our ablation study on the parameters $\alpha$ and $\lambda$ from Equations \ref{eq:2} and \ref{eq:total_loss}. Specifically, we set $\alpha = 0$ to entirely remove the Barlow Twins term from the training loss and $\lambda = 0$ to eliminate the decorrelation term.

For sensitivity analysis, we first fix $\alpha$ to its optimal value and vary $\lambda$ to assess its impact on invariance. Similarly, we fix $\lambda$ and adjust $\alpha$ to analyze sensitivity with respect to decorrelation. The results demonstrate that both invariance and decorrelation play crucial roles in learning generalizable user representations for sequential recommendation (SR).

We observe dataset-dependent optimal configurations for $\alpha$ and $\lambda$. More interestingly, tuning these parameters allows control over recommender behavior, such as the trade-off between the quality of long-tail and short-head recommendations. For instance, on Gowalla, the maximum HR@1 is achieved with $\alpha = 0.5$, while HR@10 peaks at $\alpha = 0.1$. Additionally, increasing $\alpha$ significantly improves Cov@K on YELP and Gowalla. The popularity bucket metrics further (Figure \ref{figure:ablation_alpha_buckets}) reveal that $\alpha$ can be adjusted to balance recommendation quality across different popularity segments.

\begin{figure*}[h]
    \centering
    \includegraphics[width=0.99\linewidth]{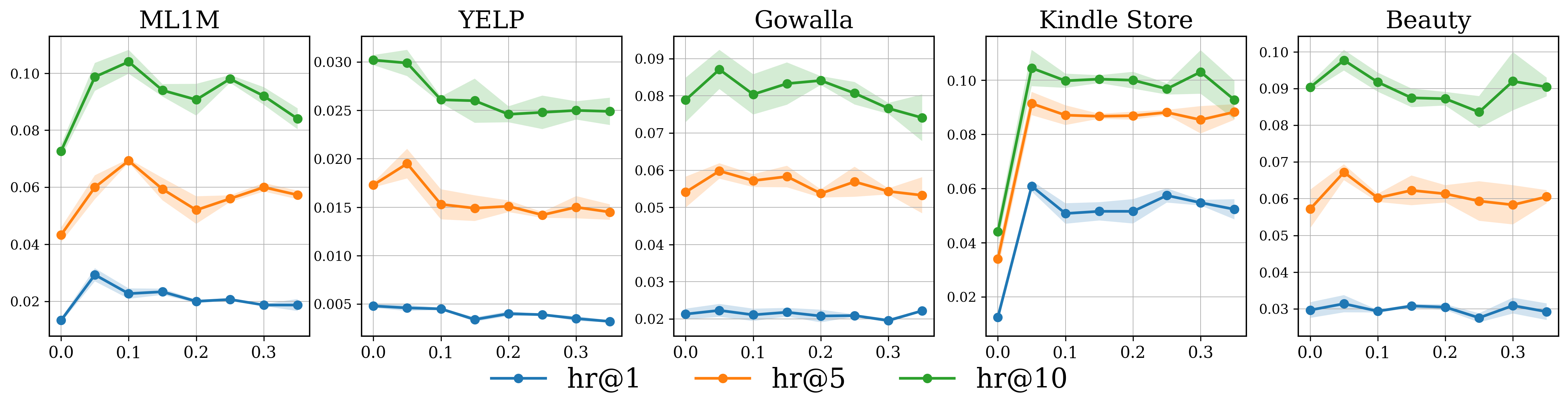}
    \includegraphics[width=0.99\linewidth]{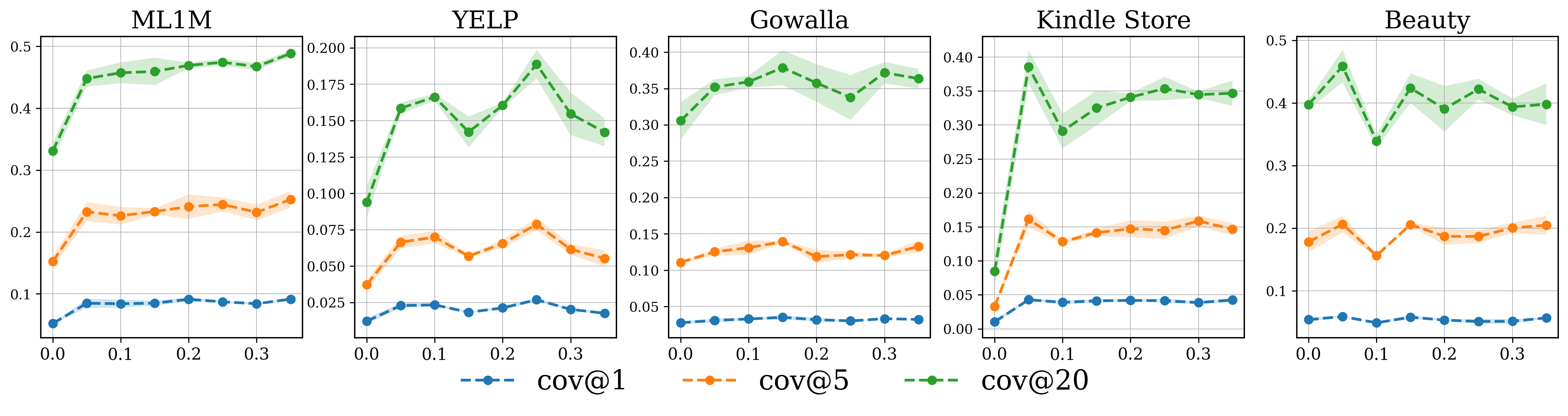}
    \caption{Sensitivity Analysis w.r.t. $ \lambda $ of hr@K and cov@K metrics }
    \label{figure:ablation_lambda_main}
\end{figure*}

\begin{figure*}[h]
    \centering
    \includegraphics[width=0.99\linewidth]{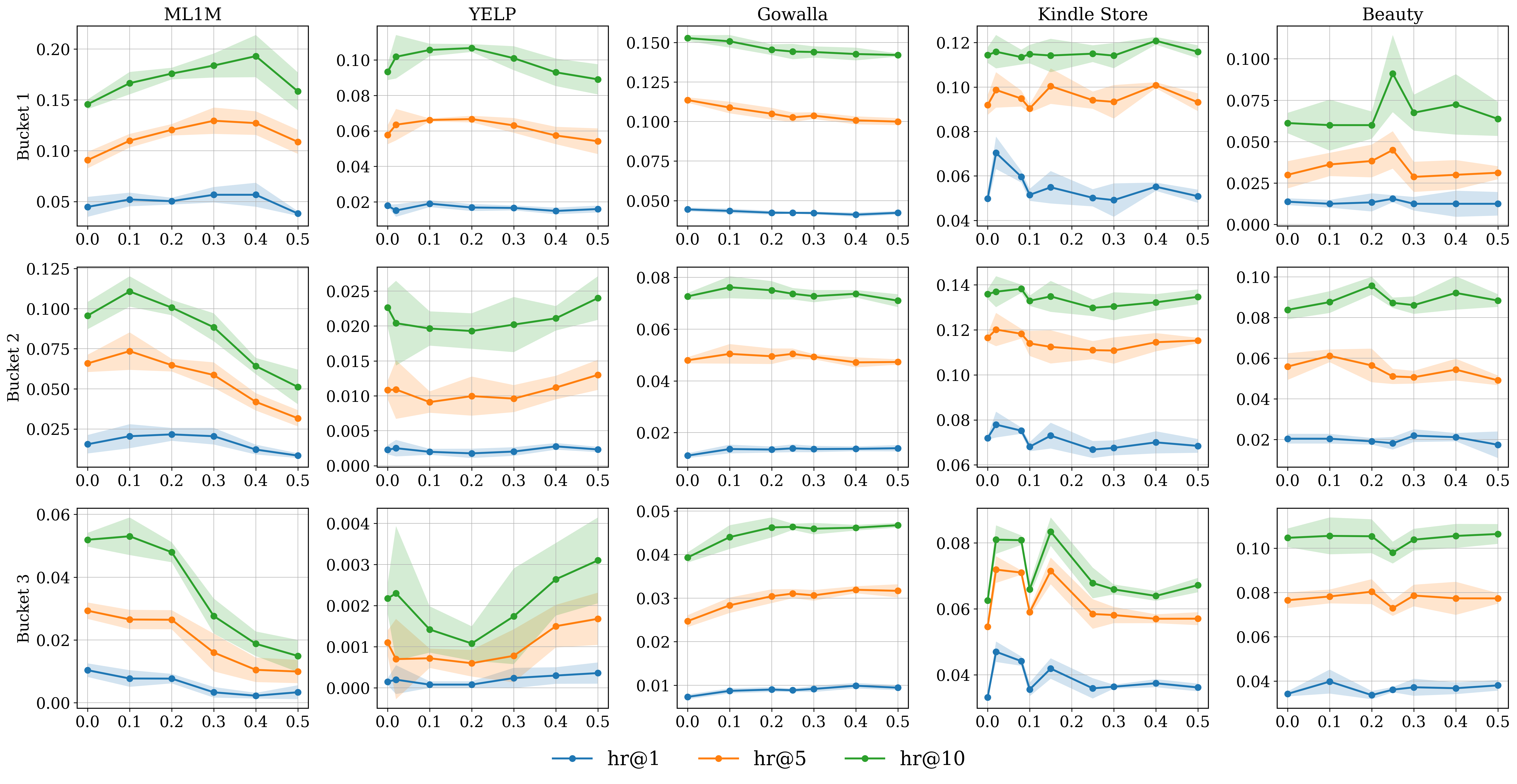}
    \caption{Sensitivity Analysis w.r.t. $ \alpha $ for recommendations over 3 item popularity buckets}
    \label{figure:ablation_alpha_buckets}
\end{figure*}

\begin{figure*}[h]
    \centering
    \includegraphics[width=0.99\linewidth]{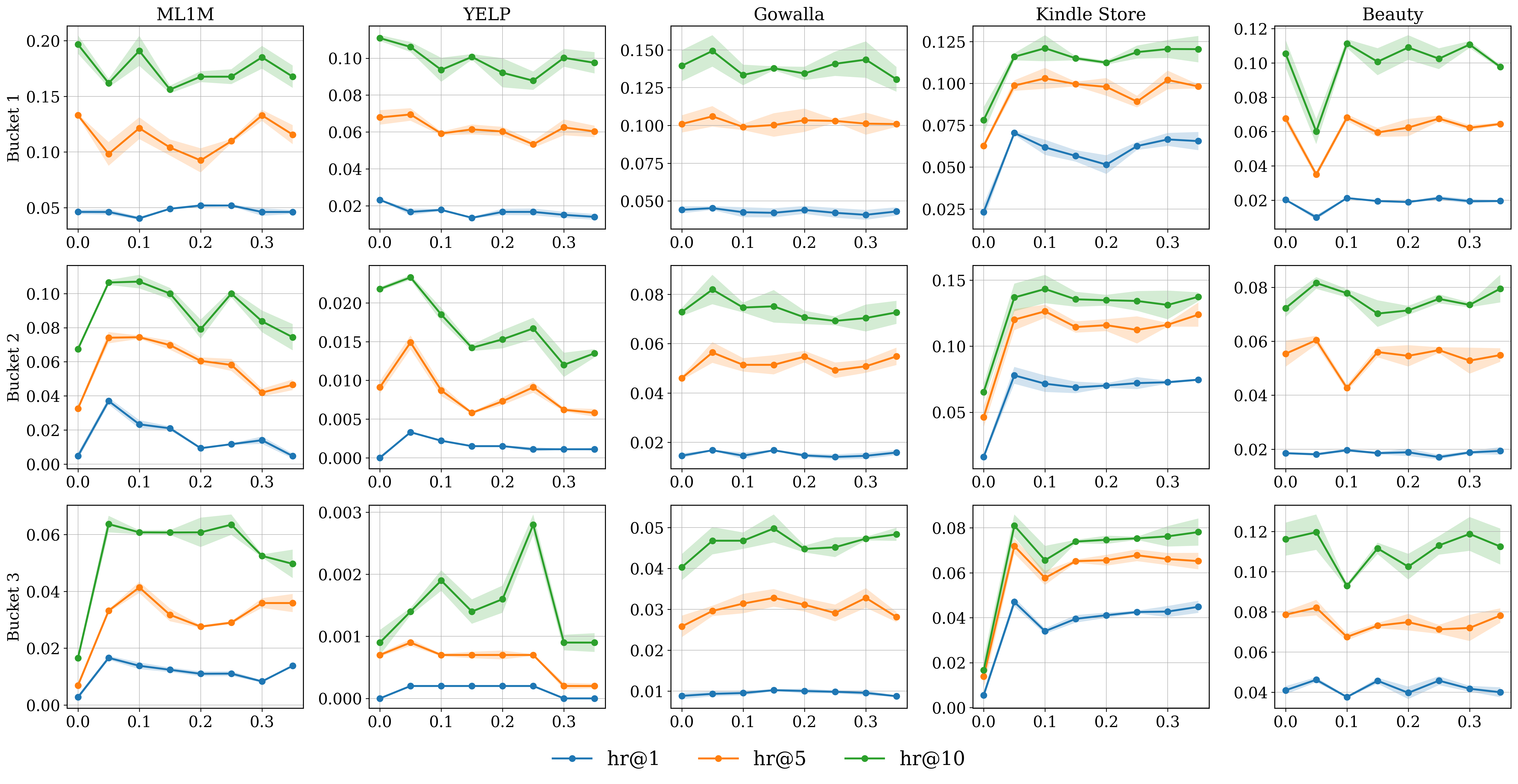}
    \caption{Sensitivity Analysis w.r.t. $ \lambda $ for recommendations over 3 item popularity buckets}
    \label{figure:ablation_lambda_buckets}
\end{figure*}

\end{document}